\documentclass[reprint,superscriptaddress,nobibnotes,amsmath,amssymb,twocolumn,prb]{revtex4-2}
\usepackage[pdftex]{graphicx}
\usepackage[english]{babel}
\usepackage{physics}
\usepackage{tabularx}
\usepackage{float}
\usepackage{amssymb}   
\usepackage{amsmath}   
\usepackage{dsfont}   
\usepackage{epsfig}
\usepackage{dcolumn}   
\usepackage{bm}        
\usepackage{amstext}
\usepackage{epsfig}
\usepackage{mdframed}
\newcommand{\be}{\begin{equation}}
\newcommand{\ee}{\end{equation}}

\newcommand{\LCPQ}{Laboratoire de Chimie et Physique Quantiques (UMR 5626), Universit\'e de Toulouse, CNRS, UPS, France}

\begin{document}
\title{ Analytic Model for the Energy Spectrum of the Anharmonic Oscillator}
\author{Michel Caffarel}
\email[]{caffarel@irsamc.ups-tlse.fr}
\affiliation{\LCPQ}
\begin{abstract}
In a recent work we have proposed an original analytic expression for the partition function of the quartic
oscillator. This partition function,
which has a simple and compact form with {\it no adjustable parameters}, 
reproduces some key mathematical properties of the exact partition function
and provides free energies accurate to a few percent over a wide range of 
temperatures and  coupling constants.
In this work, we present the derivation of the energy spectrum of this model. 
We also generalize our previous study limited to the quartic oscillator 
to the case of a general anharmonic oscillator. Numerical application for a potential of the form $V(x)=\frac{\omega^2}{2} x^2 + g x^{2m}$
show that the energy levels are obtained with
a relative error of about a few percent, a precision which we consider to be quite satisfactory 
	given the simplicity of the model, the absence of adjustable parameters, and the negligible computational cost.

\end{abstract}
\noindent
\maketitle
The one-dimensional quantum anharmonic oscillator\cite{Turbiner_2023} plays an important role in quantum mechanics as a simple yet nontrivial model 
for
describing nonlinear (anharmonic) effects. As such, it is commonly used in various scientific fields including
molecular physics (rovibrational spectra), condensed matter physics (phonons), nuclear physics (collective vibrational motions of nuclei), 
and quantum field theory ({\it e.g.}, $\phi^4$-theory, Higgs mechanism), to cite the main ones. 
No exact solution for a general anharmonic potential has been found so far, although  
some partial solutions have been developed for specific potentials, for example in the case of the so-called quasi-exactly-solvable problems;
see Turbiner\cite{Turbiner_2016} and references therein. From a numerical point of view, the simplicity of the ordinary differential equation 
to solve allows very accurate solutions.
To cite one numerical approach among many, let us mention the recently developed Lagrange Mesh method of Baye \cite{Baye_2015}
as recently implemented by del Valle\cite{del_Valle_2024}.
However, as for any physical model,
having analytical solutions -even approximate ones- is important since it may lead a deeper insight into the nature 
of the problem, reveal underlying hidden structures that may not be evident from numerical solutions and, also,
guide further analysis or generalizations. 
A great variety of approximate analytical approaches have been proposed,
making it very difficult to provide an exhaustive account of the literature. In this work, we are more specifically interested in 
evaluating the energy levels of the anharmonic oscillator. 
Among the main approaches developed for this purpose,
let us mention the semi-classical approaches, 
such as WKB\cite{Bender_1977,Kesarwani_1981} or the phase-integral method based on the generalized Bohr-Sommerfeld quantization condition (\cite{Lakshmanan_1981} and references therein) and the
methods based on the use of a variational 
approach either using parameterized excited wavefunctions\cite{Yao_1996} or path integrals\cite{,Feynman_1986,Kleinert_1992}. Other interesting
approaches include the
design of simple analytic expressions for the energy levels built from the weak- and/or strong-coupling expansions\cite{Bhattacharya_1998,Dasgupta_2007}
and methods based on the tuning of boundary conditions\cite{Leonard_2007}
or eigenvector continuation \cite{Franzke_2022}. 

The aim of this work is to present novel analytical expressions for the energy levels of the anharmonic oscillator. For that, we take advantage 
of a recently proposed partition function for the quartic oscillator,\cite{quartic}
which we extend here to an arbitrary potential.
As shown in our previous study\cite{quartic}, this partition function provides
an appealing {\it simple} and {\it physically meaningful}  model for the exact solution. Indeed, it has a simple
and compact form with {\it no adjustable parameters}, a desirable property for an analytical model. Furthermore, it reproduces 
some key mathematical properties of the exact partition function, thus supporting the idea that the important features of 
the non-trivial mathematical structure of the solution are, at least partially, accounted for. 

These properties are the following: 
i) The harmonic and classical limits are exactly recovered,
ii) the well-known divergence of the weak-coupling (Rayleigh-Schr\"odinger) expansion of the energy is reproduced.
As for the exact solution, the energy corrections are found to be rational numbers and display
a factorial-like growth in terms of the perturbational order, and
iii) the functional form of the strong-coupling expansion is recovered.

From a quantitative point of view, the free energy is found to be accurate to a few percent over a wide range of temperatures
and coupling constants. A similar precision is also obtained for 
the ground- and first-excited state energies.

To the best of our knowledge,
no partition function in closed-form proposed so far is capable of simultaneously reproducing
all these features
(see, for example, references \cite{Feynman_1965, Feynman_1986,Giachetti_1985, Buttner_1987} and a comparative study with
other models in \cite{quartic}). 

Having at our disposal a simple and physically meaningful model for the partition function, it is of interest 
to derive the full energy spectrum of the model. This is the purpose of the present work. 
As we shall see, the energy levels obtained turn out to be accurate
with a relative error of about a few percent, a precision which we consider to be quite satisfactory for such a simple model.
Of course, a higher precision can be obtained without difficulty by using some of the numerical/analytical approaches cited above.
However, we emphasize that the aim of this work is not to achieve an ultimate precision in the spectrum 
but, instead, to propose 
a simple yet faithful analytical model for the spectrum of a general anharmonic oscillator that can be readily applied 
in various scientific contexts with essentially no computational cost.\\

The Hamiltonian considered here is as follows
\be
H = -\frac{1}{2} \dv[2]{\;}{x} + V(x)
\label{H}
\ee
where $V(x)$ is a rather general potential function bounded from below and verifying $\lim_{|x| \rightarrow \infty} V(x)=+\infty$ (in other words, 
there are only discrete energies). In the numerical applications presented below, we will particularize the potential in the form
\be
V(x)=\frac{\omega^2}{2} x^2 + g x^{2m}
\label{V}
\ee
where $\omega^2$ and $g$ denote the harmonic force constant and coupling constant, respectively.\\

As a first step let us briefly recap the main steps leading to the model partition function.\\
\\
i) The partition function is first expressed as a path integral in a standard way (see, {\it e.g.}, [\onlinecite{Schulman_2005}])
\be
Z = {\rm Tr} e^{-\beta H} = \lim_{n \rightarrow \infty} Z_n
\ee
with
$$
Z_n=\qty(\frac{1}{\sqrt{2\pi \tau}})^n  \int_{-\infty}^{\infty} dx_1 ... \int_{-\infty}^{\infty} dx_n
$$
\be
e^{-\frac{1}{2\tau} \sum_{i=1}^n \qty(x_{i+1}-x_i)^2} e^{-\tau \sum_{i=1}^n V(x_i)}.
\ee
where $\tau=\frac{\beta}{n}$ is the time-step and
periodic conditions are used, $x_{n+1}=x_1$.\\

ii) Second, the short-time anharmonic contribution, $e^{-\tau V(x)}$, is approximated by a gaussian distribution centered on some position $x^*$ 
(typically, the position of the lowest minimum of the potential; here, $x^*=0$ in our applications)
with an effective frequency $\omega_g\qty(\tau)$, that is
\be
\frac{ e^{-\tau V(x)}}{\int_{-\infty}^{\infty} dx e^{-\tau V(x)}} \sim \frac{ e^{-\tau \frac{1}{2} \omega^2_g\qty(\tau) \qty(x-x^*)^2}}
{ \int_{-\infty}^{\infty} dx e^{-\tau \frac{1}{2} \omega^2_g\qty(\tau) \qty(x-x^*)^2}}.
\ee
To set the frequency $\omega_g(\tau)$ we have proposed to impose to the 
two distributions to have the same variance. In the case of the quartic oscillator treated previously,\cite{quartic} 
it leads to $\omega_g= \omega \sqrt{B\qty[\frac{4g \;}{\tau \omega^4}]}$ where $B$ is some function. 
In the more general case $V(x)=\frac{\omega^2}{2} x^2 + g x^{2m}$,
we get
\be
\omega_g\qty(\tau)= \omega \sqrt{B^{(m)}\qty[\frac{2^m g \;}{\tau^{m-1} \omega^{2m}}]}
\label{eq2}
\ee
where the parameter-free function $B^{(m)}(x)$ is given by
\be
B^{(m)}(x) = \frac{1}{2} \frac{\int_{-\infty}^{\infty} dy \; e^{-y^2 - x y^{2m}}} { \int_{-\infty}^{\infty} dy \; y^2 e^{-y^2 - x y^{2m}}}.
\label{beta}
\ee
For a general potential $V$ the formula writes 
\be
\omega_g\qty(\tau)= \sqrt{ \frac{1}{\tau} \frac{\int_{-\infty}^{\infty} dx e^{-\tau V(x)}} {\int_{-\infty}^{\infty} dx  \qty(x-x^*)^2 e^{-\tau V(x)}}}
\label{eq3}
\ee

iii) The gaussian approximation being made, the infinite-$n$ limit of $Z_n$ is no longer defined, $\lim_{n \rightarrow \infty} 
Z_n =  +\infty $. To circumvent this problem, we have proposed to introduce a 
Principle of Minimal Sensitivity (PMS) for the path integral. More precisely, we impose to the path integral 
to minimally depend on the effective frequency used in the gaussian approximation, that is,
$\frac{\partial Z_n}{\partial \omega_g(\tau)} = 0$.
For a given temperature, the PMS condition holds only
for a unique value of $n$ denoted 
as $n_c(\beta)$ [and, thus, a unique time-step, $\tau_c(\beta)=\frac{\beta}{n_c(\beta)}$].
We have then proposed to define the model partition 
function as the value of $Z_n$ at this "optimal" value of $n$, $Z \equiv Z_{n_c(\beta)}$. In particular, 
the ill-defined $n \rightarrow \infty$-limit is avoided.
The nonlinear implicit equation for $n_c(\beta)$ resulting from the PMS condition writes
\be
n_c(\beta)=\frac{ \beta \omega_g\qty[\tau_c(\beta)]}{2} {\coth{\frac{\beta \omega_g\qty[\tau_c(\beta)]}{2}}}.
\label{eq1}
\ee
iv) Finally, the analytical expression of the model partition function is given by
\be
Z= \frac{C(\beta)^{n_c(\beta)}}
{ e^{\frac{\beta \omega_g\qty[\tau_c(\beta)] }{2}}
-e^{-\frac{\beta \omega_g\qty[\tau_c(\beta)] }{2}}}
\label{Zdef}
\ee
with 
\be 
C(\beta)=  \sqrt{\frac{\omega_g\qty[\tau_c(\beta)]}{\pi \coth{\frac{\beta \omega_g\qty[\tau_c(\beta)]}{2}}} }I_g\qty(\beta) 
\label{Cbeta}
\ee
and
\be
I_g\qty(\beta)= \int_{-\infty}^{\infty} dx e^{-\tau_c(\beta) V(x)},
\label{def_I}
\ee
where the time-step $\tau_c(\beta)=\frac{\beta}{n_c(\beta)}$ is obtained from Eq.(\ref{eq1}).\\

At this point, an important remark is in order. Introducing a gaussian approximation of a non-gaussian quantity 
is a standard practice in physics (harmonic phonons, Gaussian approximation for the Ginzburg–Landau action, etc.)
In short, it is done by restricting to the second-order a Taylor expansion of some Hamiltonian or action. 
Here, our gaussian approximation is very different in nature. Instead of approximating the potential $V(x)$ 
by a quadratic potential, we approximate 
the quantity $e^{-\tau V(x)}$ by a gaussian distribution with an effective frequency which {\it depends explicitly on the time-step}. 
It is this dependence 
on the time-step that makes $Z_n$ to diverge in the large-$n$ limit, but which, after application of the PMS condition, allows to fix $n$ 
at a finite value $n_c(\beta)$ 
and, then, leads to an accurate model for the exact partition function. Approximating $V(x)$ using a quadratic
potential independent on the time-step would merely lead to the partition function of a simple harmonic oscillator and, thus, 
to a very poor model for the exact partition function.\\

{\it A. Ground-state energy.} The ground-state energy is obtained from the large-$\beta$ behavior of $Z$
\be
E_0(g) = \lim_{\beta \rightarrow \infty} -\frac{1}{\beta} \ln{Z}.
\label{e0g}
\ee
At large $\beta$, the (unique) solution $n_c(\beta)$ of Eqs.(\ref{eq2}) and (\ref{eq1}) is proportional to $\beta$ and 
given by
\be
n_c(\beta)=\frac{\beta {\bar \omega}_g}{2},
\ee
where ${\bar \omega}_g= \omega_g\qty[\tau_c(+\infty)]$ is the solution of the implicit equation given by
\be
{\bar \omega}_g= \omega \sqrt{B^{(m)}\qty( \frac{2g {\bar \omega}^{m-1}_g}{\omega^{2m}})}.
\label{implicit}
\ee
Using the zero-temperature limit of the free energy and the expression 
for the partition function we get
\be
E_0(g) = \frac{ {\bar \omega}_g }{2} \qty[ 1 -\ln{ \qty(\sqrt{\frac{{\bar \omega}_g}{\pi}} {\bar I_g})} ]
\label{E0_final}
\ee
where
\be
{\bar I}_g= \int_{-\infty}^{\infty} dx e^{ -\frac{2}{{\bar \omega}_g} V(x)}.
\label{Ig}
\ee
{\it B. Excited-state energies.} 
The exact partition function of the quartic oscillator decomposes as a discrete sum of exponentials
\be
Z= \sum_{n=0}^\infty e^{-\beta E_n}
\label{form_Z}
\ee
where $E_n$ are the excited-state energies.
As just seen, the ground-state energy $E_0$ is obtained by extracting the leading exponential component of the PF 
at large $\beta$. To get excited-state energies subleading components are to be evaluated. \\

In the following we will show that the model partition function actually does not write as a sum of simple 
exponentials, as it should be for the exact PF, but, instead, as a sum of exponentials multiplied by a polynomial term as follows
\be
Z= \sum_{n=0}^{\infty} P_n(\beta)  e^{-\beta (E_0 + n {\bar \omega}_g)}
\label{Z_improved}
\ee
where $P_n(\beta)$ is a polynomial of degree $n$ in $\beta$ with $P_n(0)=1$.
The partition function being no longer expressed as a sum of simple exponentials, 
the definition of what is meant by excited-state energies becomes problematic.
Comparisons with the "exact" numerical spectrum show that
defining the excited energies as the exponents of the exponential contributions, that is,
\be
E_n= E_0 + n {\bar \omega}_g
\ee
gives a very poor approximation of the exact spectrum. 
Note that this is not surprising since, in such a case, the energy 
differences between successive states would remain constant, a property which is clearly 
wrong for the exact spectrum. In sharp contrast, we have found that modifying the partition function by 
exponentiating the linear contribution of the polynomials and incorporating it into the exponential part leads
to remarkably good energy levels. Precisely, we propose to replace the polynomial 
\be
P_n(\beta) = 1 + P_{n1} \beta + ... + P_{nn} \beta^n 
\label{approx}
\ee
by $e^{\beta P_{n1}}$ and to define the excited-state energies of our model as
\be
E_n= E_0 + n {\bar \omega}_g - P_{n1}.
\label{En}
\ee
Let us emphasize that this replacement should not be considered
as a quantitative approximation, but rather, as a "minimal"
modification of the model to impose to the partition function to have the exact expansion, Eq.(\ref{form_Z}).
Unfortunately, we have not been able to 
understand why this simple additional prescription to our model is so effective in giving accurate energy 
levels (see, the figures to follow). However, it should be considered as a salient result of this work.

To derive the functional form of the partition function, Eq.(\ref{Z_improved}), we first need to expand 
the effective frequency $\omega_g\qty[\tau_c(\beta)]$ as a power series of exponential-like contributions.

Let us first consider the case $\omega \ne 0$.
Using Eqs.(\ref{eq2}) and (\ref{eq1}), and introducing the
variable $y$ defined as
\be
y=e^{-\beta \omega_g}
\ee
the implicit equation obeyed by $\omega_g$ is rewritten as
\be
\omega_g= \omega 
\sqrt{ B\qty[ \frac{2g}{\omega^{2m}}  \omega_g^{m-1}\qty( 1 + 2 \sum_{n=1}^{\infty} y^n)^{m-1}]}
\label{eqomeg}
\ee
where the coth function has been expanded in power series of $y$. Note that the superscript $(m)$ has been removed from $B^{(m)}(x)$ to 
simplify the notation. This will also be the case in the following for most quantities depending on $m$ when no confusion is possible.
The function $B$ being infinitely differentiable, 
$\omega_g$ can be expanded in powers of $y$
\be
\omega_g=  {\bar \omega}_g + \sum_{n=1}^\infty \omega_n y^n.
\ee
Remark that the equation obeyed by $\omega_g$ has no {\it explicit} dependence on $\beta$ 
(the dependence on the temperature is only 
through the variable $y$). Accordingly, the coefficients $\omega_n$ are independent of $\beta$. 
Their evaluation is done i) by introducing the expansion of $\omega_g$
in Eq.(\ref{eqomeg}), ii) by Taylor-expanding the function 
$B$ at $x_0=\frac{2g}{\omega^{2m}} {\bar \omega}^{m-1}_g$, and, finally, by identifying the contributions corresponding to 
a given power of $y$. After some algebra, an explicit expression for the $\omega_n$'s can be derived.
For example, the first coefficient, $\omega_1$'s is given by
\be
\omega_1= {\bar \omega}_g  \frac{\frac{B_1}{B_0} x_0 (m-1)}{1-\frac{\frac{B_1}{B_0} x_0}{2} (m-1) }
\ee
where $B_0=B^{(m)}(x_0)$ and $B_1=\frac{dB^{(m)}}{dx}(x_0)$. For $n \ge 2$, $\omega_n$ can be expressed as a 
function of the preceding coefficients, $\omega_p$ with $ 1 \le p \le n-1$. The explicit formula is given in Appendix \ref{appendix_A}.

When $\omega=0$ the coefficients $\omega_k$ are more easily derived. From the general expression of the effective frequency, Eq.(\ref{eq3}),
we get $\omega_g(\tau)=c_0 \qty(\frac{g}{\tau})^{\frac{1}{4}}$ with $c_0=\sqrt{ \frac{ \int dx e^{-x^4}}{\int dx x^2 e^{-x^4}}}$.
Using the PMS condition, Eq.(\ref{eq1}), we obtain
\be
\omega_k = {\bar \omega}_g \sum_{l=1}^k 2^l \binom{\frac{1}{3}}{l} \binom{k-1}{l-1} \;\;\; k \ge 1
\label{eq27}
\ee
with
\be
{\bar \omega}_g= c^{\frac{4}{3}}_0 \qty(\frac{g}{2})^{\frac{1}{3}} .
\label{eq28}
\ee
In this $\omega=0$-case, the ground-state energy can be explicitly written as
\be
E_0= \qty[ \frac{\Gamma\qty(\frac{5}{4})}{\Gamma\qty(\frac{3}{4})}]^{\frac{2}{3}}
 \qty(1-\log{  \sqrt{  \frac{ \qty[2\Gamma\qty(\frac{5}{4})]^3 }{\pi \Gamma\qty(\frac{3}{4})  }   }})  g^\frac{1}{3}.
\ee

Now, to proceed, let us introduce the following new variable $y_0$ 
\be
y_0 = e^{-\beta {\bar \omega}_g}.
\ee
The next step consists in expressing the partition function as a power series in $y_0$, thus leading to Eq.(\ref{Z_improved}). 
For that, we first need to derive the expansion of the effective frequency in terms of $y_0$. We have
\be
\omega_g = {\bar \omega}_g  + \sum_{n=1}^\infty \omega_n y^n_0 e^{-\beta n \qty[\omega_g-{{\bar \omega}_g}]}.
\label{expan}
\ee
By expanding the exponential in power series and by simple inspection, the form of the solution is 
\be
\omega_g = {\bar \omega}_g + \sum_{n=1}^\infty Q_n(\beta) y^n_0
\label{expansion_omegag}
\ee
where $Q_n(\beta)$ are polynomials of degree $n-1$ in $\beta$.
The polynomials 
can be evaluated by differentiation
\be
Q_n(\beta) =\frac{1}{n!} \frac{ \partial^n \omega_g}{\partial y^n_0}(y_0=0).
\ee
Using Eq.(\ref{expan}) and the Leibniz formula for derivatives, the $n$-th derivative of $\omega_g$ writes
\be
\frac{1}{n!} \frac{ \partial^n \omega_g}{\partial y^n_0}(y_0=0) = \sum_{k=1}^n \omega_k  \frac{1}{(n-k)!}
\frac{\partial^{n-k}} {\partial y^{n-k}_0} \qty[ e^{-\beta k (\omega_g -{\bar \omega}_g)}](y_0=0).
\ee
To proceed we make use of the Faà di Bruno formula
\be
\frac{\partial^n} {\partial x^n} e^{f(x)} = e^{f(x)} \sum^\prime_{m_1, m_2 \;..., m_n} \frac{n!} { m_1!m_2! \; ...\; m_n!} \prod_{j=1}^n \qty[ \frac{ f^{(j)}(x)}{j!}]^{m_j}   
\ee
where the prime on the sum indicates summing with the constraint
\be
1 m_1 + 2 m_2 +\; .... \; n m_n = n.
\ee
We then have
$$
\frac{1}{n!} \frac{ \partial^n \omega_g(\beta)}{\partial y^n_0} (y_0=0)
= \sum_{k=1}^n \omega_k  \frac{1}{(n-k)!}
$$
$$
\times \sum^\prime_{m_1, m_2 \;..., m_{n-k}} \frac{(n-k)!}{m_1!m_2! \; ...\; m_{n-k}!}
\prod_{j=1}^{n-k} \qty[ \frac{ (-k \beta)(\omega_g)^{(j)}(x)}{j!}]^{m_j}(y_0=0)
$$
and, finally 
$$
Q_n(\beta) = \omega_n+ \sum_{k=1}^{n-1} \omega_k 
$$
$$
\times \sum^\prime_{m_1, m_2 \;..., m_{n-k}} 
\frac{ \qty[-k\beta Q_1(\beta)    ]^{m_1}   } {m_1!    } ... 
\frac{\qty[-k\beta Q_{n-k}(\beta) ]^{m_{n-k}}} {m_{n-k}!},
$$
a relation which allows to evaluate iteratively the polynomials $Q_n(\beta)$.
From this expression, we see that $Q_n(\beta)$ are indeed polynomials of degree $n-1$ as stated above.
Let us give the first five polynomials
$$
Q_1(\beta)= \omega_1
$$
$$
Q_2(\beta)= \omega_2 - \beta \omega^2_1
$$
$$
Q_3(\beta)= \omega_3 - 3 \beta \omega_1 \omega_2 + \frac{3}{2} \beta^2  \omega^3_1
$$
$$
Q_4(\beta)= \omega_4 - 2 \beta ( 2 \omega_1 \omega_3 + \omega^2_2 ) + 8 \beta^2 \omega_2 \omega^2_1 -\frac{8}{3} \omega^4_1 \beta^3
$$
$$
Q_5(\beta)= \omega_5 - 5 \beta ( \omega_2 \omega_3 + \omega_1 \omega_4) + \frac{25}{2} \beta^2 (\omega_3 \omega^2_1 +
\omega_1 \omega^2_2)
$$
$$
-\frac{125 \beta^3}{6} \omega^3_1 \omega_2  +
\frac{125}{24} \omega^5_1 \beta^4
$$

We are now ready to derive the expansion of the partition function in the variable $y_0$. 
It is convenient to introduce the following three quantities
\be
\Delta \omega_g \equiv \omega_g -{\bar \omega}_g,
\ee
\be
\Delta T \equiv \ln{C(\beta)} - \ln{\sqrt{\frac{{\bar \omega}_g}{\pi}} {\bar I}_g},
\label{def_DT}
\ee
and
\be
\Delta R \equiv \frac{1}{2} (\coth{ \frac{\beta \omega_g}{2}} - 1).
\ee
The partition function can be written as
\be
Z = e^{-\beta E_0} P.
\label{Z1}
\ee
Simple algebra shows that $P$ can be expressed as 
\be
P= e^{\beta S} \qty[ 1 + \Delta R]
\label{Pbeta}
\ee
where
$$
S= c_1 \Delta \omega_g  + c_2 \Delta T  + c_3 \Delta R
$$
\be
+  c_4 \Delta \omega_g  \Delta T + c_5 \Delta \omega_g \Delta R+ c_6 \Delta T \Delta R +
c_7  \Delta \omega_g  \Delta T \Delta R
\label{defS}
\ee
with the following coefficients
$$
c_1= \frac{1}{2} \qty(\ln{\sqrt{\frac{{\bar \omega}_g}{\pi}} {\bar I}_g}-1) 
$$
$$
c_2 = \frac{{\bar \omega}_g}{2} 
$$
$$
c_3 = {\bar \omega}_g \ln{ 
\sqrt{\frac{{\bar \omega}_g}{\pi}} {\bar I}_g}
$$
$$
c_4=  \frac{1}{2}   
$$
$$
c_5 = \ln{\sqrt{\frac{{\bar \omega}_g}{\pi}} {\bar I}_g}
$$
$$
c_6 = {\bar \omega}_g 
$$
$$
c_7=1
$$
Let us introduce the following form for the expansion of a quantity $X$
\be
X = \sum_{n=1}^{\infty} X_n(\beta) y_0^n
\label{rep_X}
\ee
where $X_n$ is a polynomial of order $n-1$ in $\beta$. Note that this form is stable by multiplication by a scalar, 
addition, multiplication, and exponentiation. We have already seen that $\Delta \omega_g$ admits 
such a representation.
From the relation
\be
y =y_0 e^{-\beta \Delta \omega_g}
\ee
we see that it is also the case for $y$. The quantities 
$\Delta T$ and $\Delta R$ can be written as a function of the variable $y$ only. By expanding these quantities in terms of $y$ 
we thus find that they also both admit this representation.
From (\ref{defS}) it is also true for $S$. Finally, 
the partition function, Eqs.(\ref{Z1},\ref{Pbeta}), can be written as
\be
Z= e^{-\beta E_0} \qty[ 1 + \sum_{n=1}^\infty P_n(\beta) y_0^n ]
\ee
where $P_n(\beta)$ is a polynomial of order $n$ in $\beta$. This is the form 
given in Eq.(\ref{Z_improved}).\\

To get the excited-state energies the quantities $P_{n1}$ need to be evaluated, see Eq.(\ref{En}).
Using the formula
$$
P_{n1} = \frac{1}{n!} \frac{ \partial^{n+1} P(\beta)}{\partial \beta \partial y^n_0}(\beta=0,y_0=0)
$$
we get
\be
P_{n1} = S_{n0} + \Delta R_{n1} + \sum_{k+l=n \;\; k \ge 1 \; l \ge 1} S_{k0} \Delta R_{l0}.
\label{Pn1}
\ee
where $S_{nk}$ and $\Delta_{nk}$ are the coefficients of the $n$-th order polynomial in the representation, Eq.(\ref{rep_X}),
for $S$ and $\Delta R$, respectively.
Using Eq.(\ref{defS}) and $Q_{n0}= \omega_n$ we have
$$
S_{n0}= c_1 \omega_n + c_2 \Delta T_{n0} + c_3 \Delta R_{n0}
$$
$$
+c_4\sum_{k_1+k_2=n} \omega_{k_1} \Delta T_{k_20} +c_5\sum_{k_1+k_2=n} \omega_{k1} \Delta R_{k_20} 
$$
$$
+
c_6 \sum_{k_1+k_2=n} \Delta T_{k_10} \Delta R_{k_20}
$$
\be
+ c_7  \sum_{k_1+k_2=n} \qty( \sum_{k_3+k_4=k_1} \omega_{k_3} \Delta T_{k_40}) \Delta R_{k_20}
\label{Sn0}
\ee
where $\Delta T_{nk}$ are the coefficients of the polynomial for $\Delta T$. 
In each sum above, $k_i \ge 1$.
Let us evaluate the quantities $\Delta R_{n0}$, $\Delta R_{n1}$, and $\Delta T_{n0}$.\\

Starting from 
$$
\Delta R = \sum_{n=1}^\infty y^n_0 e^{-\beta n \Delta \omega_g}
$$
we have
$$
\Delta R= \sum_{n=1}^\infty y^n_0 \sum_{k=1}^\infty \frac{ (-\beta n)^k}{k!} \qty(\sum_{l=1}^\infty Q_l(\beta) y^l_0)^k
= \sum_{n=1}^\infty y^n_0 \sum_{k=1} \alpha_k y^k_0
$$
with
$$
\alpha_k=\sum_{l=1}^k \frac{ (-\beta n)^l}{l!}  \sum_{k_1+ ...+ k_l=k}  Q_{k_1}(\beta) ...  Q_{k_l}(\beta)
$$
which gives
\be
\Delta R_{n0}=1, \; \Delta R_{11}=0, \; {\rm and} \;\Delta R_{n1} = -\sum_{k=1}^{n-1} k \; \omega_{n-k}  \; {\rm for} \; n \ge 2
\label{DR_n1}
\ee

Let us now calculate $\Delta T_{n0}$. $\Delta T$, as defined by Eq.(\ref{def_DT}), is decomposed as
$$
\Delta T = \Delta A + \Delta B
$$
with
$$
\Delta A= \ln{ \sqrt{\frac{\omega_g(\beta)}{\pi \coth{\frac{\beta \omega_g\qty[\tau_c(\beta)] }{2}}  }} }
-\ln{ \sqrt{\frac{{\bar \omega}_g}{\pi}} }
$$
and
$$\Delta B= \ln{ I_g(\beta)} - \ln{{\bar I}_g} $$
Let us begin with $\Delta A$ which can be written as
$$
\Delta A= \frac{1}{2} \ln{ \qty[ \frac{ 1 + \sum_{n=1}^\infty \frac{Q_n(\beta)}{{\bar \omega}_g} y^n_0 }
{ 1 + \sum_{n=1}^\infty 2 \Delta R_n(\beta) y^n_0 } ]}.
$$
Expanding the two logarithmic terms, we get after some algebra
$$
\Delta A_{n0} =\frac{1}{2} \sum_{k=1}^n \frac{(-1)^{k-1}} {k} 
\sum_{l_1 + ... +l_k=n} 
\qty[ \qty( \frac{\omega_{l_1}}{{\bar \omega}_g} ) ...
\qty( \frac{\omega_{l_k}}{{\bar \omega}_g} )  - 2^k].
$$
\\
We are now left wit the calculation of 
$\Delta B_{n0}$. We have
$$
I_g(\beta)= \sum_{n=0}^\infty \frac{(-1)^n {\bar I}_n}{n!} \Delta\tau^n_c
$$
with
$$
\Delta \tau_c \equiv \tau_c-{\bar \tau}_c
$$
where ${\bar \tau}_c=\tau_c(\infty)=\frac{2}{{\bar \omega}_g}$ and
$$
{\bar I}_n  \equiv \int_{-\infty}^{\infty} dx x^n e^{-{\bar \tau}_c V(x)}.
$$
After some algebra
$$
\Delta \tau_c= \frac{2}{{\bar \omega}_g} 
\sum_{n=1}^\infty (-1)^n \qty[ \sum_{k=1}^{\infty} X_k(\beta) y^k_0 ]^n
$$
with
$$
X_k(\beta)= 2 \Delta R_k(\beta) + \frac{Q_k(\beta)}{{\bar \omega}_g} + \frac{2}{{\bar \omega}_g} \sum_{l+m=k \; l\ge 1 \; m\ge 1} 
Q_l(\beta) Q_m(\beta).
$$
We then have
$$
\Delta \tau_c =  \sum_{n=1}^\infty Y_n(\beta) y^n_0
$$
with
$$
Y_{n} =  \frac{2}{{\bar \omega}_g}\sum_{k=1}^n (-1)^k  \sum_{l_1+ ...+l_k=n} X_{l_1}... X_{l_k}
$$
Then
$$
I_g=  {\bar I}_g+ \sum_{n=1}^\infty  \frac{(-1)^n {\bar I}_n}{n!} \qty(\sum_{k=1}^\infty Y_k(\beta) y^k_0)^n
$$
$$
I_g= {\bar I}_g + \sum_{n=1}^\infty Z_n(\beta)  y^n_0
$$
$$
Z_{n} = \sum_{k=1}^n (-1)^k \frac{{\bar I}_k}{k!}  \sum_{l_1+ ...+l_k=n} Y_{l_1}... Y_{l_k}
$$
Thus
$$
\ln{I_g} = \ln{{\bar I}_g} + \ln{\qty( 1 + \sum_{n=1}^\infty \frac{ Z_n}{{\bar I}_g} y^n_0)}.
$$
Finally,
$$
\ln{I_g} = \ln{{\bar I}_g} + \sum_{n=1}^\infty \Delta B_n(\beta)  y^n_0
$$
with
$$
\Delta B_{n}= \sum_{k=1}^n \frac{(-1)^k}{k {\bar I}^k_g} \sum_{l_1+ ... +l_k=n} Z_{l_1}... Z_{l_k}.
$$
The quantities $\Delta B_{n0}$ are then evaluated by taking $\beta=0$ in the preceding expressions.\\

\begin{figure}[h!]
\centering
\includegraphics[scale=0.40]{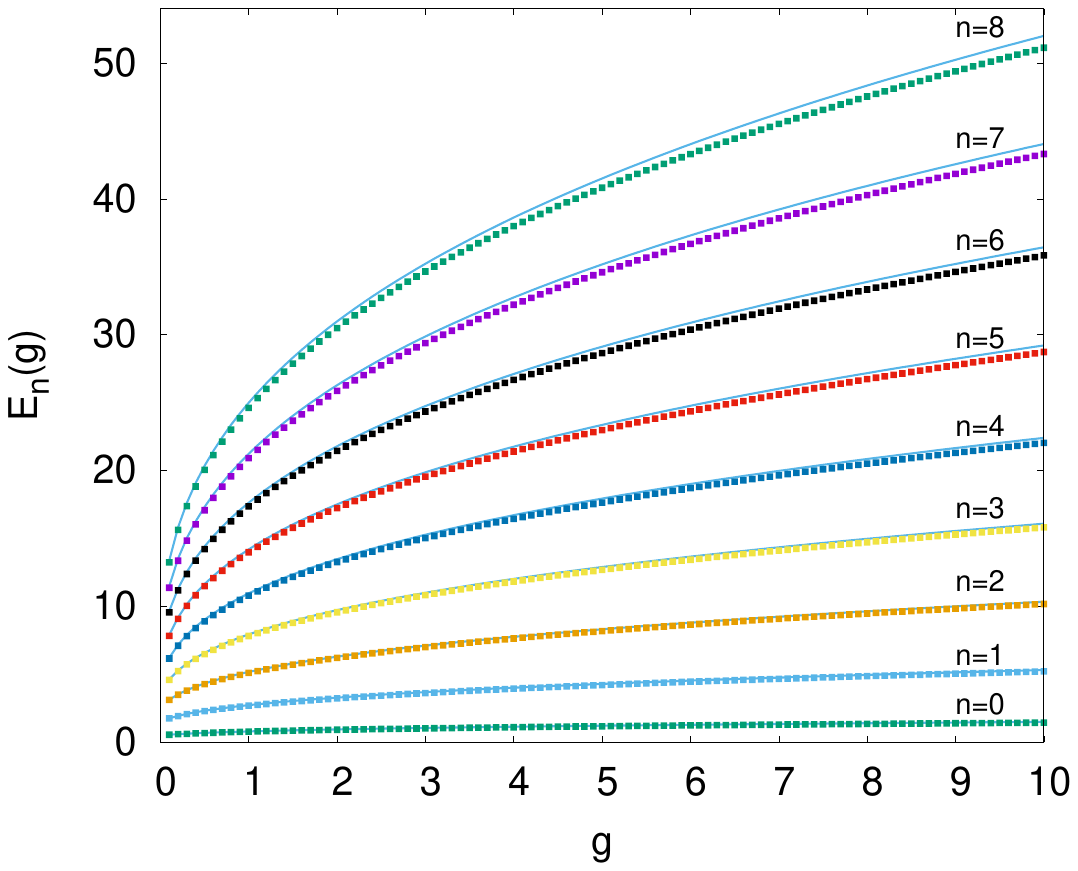}
\caption{Energy levels of the quartic oscillator as a function of $g$ for $n=0$ to $n=8$. 
Exact results given by the solid lines. Harmonic frequency, $\omega=1$}
\label{fig1}
\end{figure}

Figure \ref{fig1} shows the ground-state energy and the first eight excited state energies 
of the quartic oscillator with $\omega=1$ versus the coupling constant $g$.
The "exact" numerical energies, obtained by diagonalization of the Hamiltonian in a sufficiently 
large Gaussian basis set, are shown as solid lines.
At the scale of the figure, the energy levels and their overall behavior as functions of
$g$ and $n$ are well reproduced and appear satisfactory.
Quantitative results are presented in Table \ref{tab1}, which reports the relative errors $\epsilon$ in the computed energies.
We present results for three different potentials: the first two are the quartic ($m=2$, $\omega=1$) and sextic 
($m=3$, $\omega=1$) anharmonic potentials, respectively.
The third is the "pure" quartic potential, defined by the absence of a quadratic term ($\omega=0$).
Results are given for five values of $g$ ($g=0.1,1,10,40$, and $400$), spanning the weak- to strong-coupling regimes.
A first remark is that relative errors are all negative. In other words, the computed energies are always smaller than the exact ones. 
Unfortunately, we were not able to understand the origin of this interesting observed property. A second remark is that relative errors 
are all of the order of a few percent (up to 10\% in the worst case of the sextic potential at large $g$). 
We consider this level of accuracy as quite satisfactory 
in view of the simplicity of the model. Let us insist on the fact that the model partition function 
has a particularly simple and compact form, Eq.(\ref{Zdef}) 
with Eqs.(\ref{Cbeta}) and (\ref{def_I}); and, most importantly, no adjustable parameters. In addition, 
the computational cost for calculating the energies is negligible.
Another remark is that for each potential the relative error is maximal for the two lowest energies. Quite remarkably, the errors 
on the higher energies ($n \ge 2$) are almost constant and, also, nearly independent on the value of $g$. 
This is a quite interesting feature of the model.
The comparison between the three different potentials is instructive. As expected, the accuracy reached for the sextic oscillator is 
inferior to that obtained for
the quartic oscillator, a consequence of the  greater anharmonicity of the former potential.
This is also the case at small values of $g$ for the pure quartic oscillator in which the quadratic term has been removed. However, for large $g$'s
the errors become nearly identical.
A striking feature of the energies of the pure quartic potential is that the relative errors are independent on $g$ 
(actually, all digits of the errors are identical for the different $g$'s, a result
not shown here). This result is explained as follows. By a simple rescaling of the Schr\"odinger equation,
the exact energies $E_n(g)$ of the pure quartic oscillator can be shown to scale as
$ g^{\frac{1}{3}}$. Quite satisfactorily, it is also the case for our model. This can be proved by 
noting that ${\bar \omega_g}$, Eq.(\ref{eq28}), and the $\omega_k$'s, Eq.(\ref{eq27}) scale also as $g^{\frac{1}{3}}$
and, then, by invoking the series of equations, (\ref{En}),(\ref{Pn1}),(\ref{Sn0}) and (\ref{DR_n1}). The exact and model 
energies having the same scaling in $g$, the relative errors are thus independent on $g$.

\begin{table}[htb]
\centering
	\caption{Relative errors $\epsilon$ (in $\%$) on the energy levels $E_n$ for different values of $g$ and potentials
	 $V_{[a,n]}(x)=a x^2 + g x^n$; $\epsilon=(E_{\text model}-E_{ex})/E_{\text ex}$.}
        \label{tab1}
\begin{tabular}{lccccccccc}
	                & E$_0$ &E$_1$ &E$_2$ &E$_3$ &E$_4$ &E$_5$ &E$_6$ &E$_7$ &E$_8$ \\
g=0.1\\
\;\;	$V_{[\frac{1}{2},4]}$& -0.5 &\;-0.4&\; -0.5&\; -0.6&\; -0.7&\; -0.8&\; -0.8&\; -0.9& -0.9\\
\;\;	$V_{[\frac{1}{2},6]}$& -2.5 &\;-2.6&\; -2.7&\; -2.9&\; -3.1&\; -3.2&\; -3.3&\; -3.4& -3.4\\
\;\;	$V_{[0,4]}$          & -4.2 &\;-2.0&\; -1.6&\; -1.7&\; -1.7&\; -1.7&\; -1.7&\; -1.7& -1.7\\
g=1\\                                   
\;\;  $V_{[\frac{1}{2},4]}$& -2.3 &\;-1.4&\; -1.3&\; -1.4&\; -1.4&\; -1.4&\; -1.4&\; -1.5& -1.5\\
\;\;  $V_{[\frac{1}{2},6]}$& -6.3 &\;-4.6&\; -3.7&\; -3.7&\; -3.7&\; -3.7&\; -3.8&\; -3.8& -3.8\\
\;\;  $V_{[0,4]}$          & -4.2 &\;-2.0&\; -1.6&\; -1.7&\; -1.7&\; -1.7&\; -1.7&\; -1.7& -1.7\\
g=10\\                                  
\;\;  $V_{[\frac{1}{2},4]}$& -3.7 &\;-1.9&\; -1.6&\; -1.6&\; -1.6&\; -1.6&\; -1.6&\; -1.6& -1.6\\
\;\;  $V_{[\frac{1}{2},6]}$& -9.0 &\;-5.4&\; -4.0&\; -4.0&\; -4.0&\; -3.9&\; -3.9&\; -3.9& -3.9\\
\;\;  $V_{[0,4]}$          & -4.2 &\;-2.0&\; -1.6&\; -1.7&\; -1.7&\; -1.7&\; -1.7&\; -1.7& -1.7\\
g=40\\                                  
\;\;  $V_{[\frac{1}{2},4]}$& -4.0 &\;-2.0&\; -1.6&\; -1.7&\; -1.6&\; -1.6&\; -1.6&\; -1.6& -1.6\\
\;\;  $V_{[\frac{1}{2},6]}$& -9.0 &\;-5.5&\; -4.1&\; -4.1&\; -4.1&\; -4.0&\; -4.0&\; -4.0& -4.0\\
\;\;  $V_{[0,4]}$          & -4.2 &\;-2.0&\; -1.6&\; -1.7&\; -1.7&\; -1.7&\; -1.7&\; -1.7& -1.7\\
g=400\\                                 
\;\;  $V_{[\frac{1}{2},4]}$& -4.2 &\;-2.0&\; -1.6&\; -1.7&\; -1.7&\; -1.7&\; -1.7&\; -1.7& -1.7\\
\;\;  $V_{[\frac{1}{2},6]}$& -10. &\;-5.7&\; -4.2&\; -4.1&\; -4.0&\; -4.0&\; -4.0&\; -4.0& -4.0\\
\;\;  $V_{[0,4]}$          & -4.2 &\;-2.0&\;-1.6 &\; -1.7&\; -1.7&\; -1.7&\; -1.7&\; -1.7& -1.7\\
\hline
\end{tabular}
\end{table}

\acknowledgments{I would like to thank the Centre National de la Recherche Scientifique (CNRS) for its continued support.
I also acknowledge funding from the European Research Council (ERC) under the European Union's Horizon 2020 research and innovation programme (Grant agreement No.~863481).
}
\appendix
\newpage

\section{Formula for evaluating $\omega_n$'s}
\label{appendix_A}
In this appendix, we give the explicit formulas for the coefficients of the expansion of $\omega_g$ 
in powers of $y$ written as
\be
\omega_g=  {\bar \omega}_g + \sum_{n=1}^\infty \omega_n y^n.
\ee
Note that to simplify the notation the dependence on $m$ of the various quantities considered here will be 
omitted.

Let us define now the following quantities
\be
B_n \equiv \frac{d^nB^{(m)}}{dx^n}(x_0) 
\ee
\be
x_0=\frac{2g}{\omega^{2m}} {\bar \omega}^{m-1}_g
\ee
\be
\alpha_n= \frac{1}{n!} \frac{B_n}{B_0} x_0^n
\ee
and
\be
\beta_n = \frac{\frac{1}{2}\qty(\frac{1}{2}-1)...\qty(\frac{1}{2}-\qty(n-1))}{n!}
\ee
${\bar \omega_g}$ is given by
\be
{\bar \omega_g}= \omega \sqrt{ B_0}
\ee
The first coefficient $\omega_1$ is given by
\be
\omega_1 = A \alpha_1 (m-1) 
\ee
with
\be
A =\frac{{\bar \omega_g}} { 1- \frac{ \alpha_1 (m-1) }{2 }  }
\ee
For $n \ge 2$ the $\omega_n$'s are function of the previous coefficients $\omega_p$ with $1 \le p \le n-1$ as follows
\be
\omega_n=A \qty[ \alpha_1 (m-1) S_n +\frac{1}{2} \alpha_1 T_n + U_n + V_n]
\ee
where
\be
S_n = 1 +\sum_{k+l=n \; k \ge 1 \; l \ge 1} \frac{\omega_k}{{\bar \omega_g}}
\ee
\be
T_n = \sum_{k=2}^{m-1} \binom{m-1}{k} \sum_{l_1+...+ l_{k}=n \; l_i \ge 1} a_{l_1} ... a_{l_k} 
\ee
with
\be
a_n = \frac{\omega_n}{{\bar \omega_g}} + 2 S_{n}   \;\; n \ge 1
\ee
\be
U_n = \frac{1}{2} \sum_{k=2}^n \alpha_k \sum_{l_1+...+l_k=n \;\; l_i \ge 1 } b_{l_1} ... b_{l_k}
\ee
with
\be
b_n = (m-1) a_n + T_n  \;\; n \ge 1
\ee
and
\be
V_n = \sum_{k=2}^n \beta_k \sum_{l_1+...+l_k=n \;\; l_i \ge 1} c_{l_1} ... c_{l_k}
\ee
with
\be
c_n =\alpha_1 b_n + 2 U_n  \;\; n \ge 1
\ee
When no index fulfills the constraint in the sum, the corresponding quantity is equal to zero. Here, it means that
$S_1=1$, $T_1=0$, $U_1=0$, $V_1=0$, and $T_n=0$ for $m=2$.


\end{document}